\documentclass[a4paper,papersize]{jps-cp}
\usepackage{txfonts} 

\title{Impact of Rotation on Quark--Hadron Hybrid Stars}

\author{Tomoki \textsc{Endo}}

\inst{Institute of Education, Osaka Sangyo University, 3-1-1 Nakagaito, Daito, Osaka 574-8530, Japan}

\email{endo@edo.osaka-sandai.ac.jp}

\recdate{June 16, 2017}

\abst{Many recent observations give restrictions to the equation of state (EOS) for high-density matter. Theoretical studies are needed to try to elucidate these EOSs at high density and/or temperature. With the many known rapidly rotating neutron stars, e.g., pulsars, several theoretical studies have tried to take into account the effects of rotation. In our study of these systems, we find that one of our EOSs is consistent with recent observation, whereas the other is inconsistent.}

\kword{neutron stars, rotation effect}

\begin{document}
\maketitle

\section{Introduction}

It is widely stated that neutron stars are astronomical objects having high-density interiors and various hadronic-matter phases are expected to appear in their inner cores. Several studies show that the interior structures affect the macroscopic features and behavior of such stars.
These structures strongly depend on the equation of state (EOS) for high-density matter. However, it remains unclear what the EOS, consistent with recent observations  \cite{demo, anto}, is. Therefore, we are forced to speculate on the EOS in the theoretical study. The Tolman--Oppenheimer--Volkoff (TOV) equation  \cite{shapiro} assumes that the star is spherical. However, since neutron stars are observed as pulsars, which are known to spin rapidly, it should be considered to be elliptical rather than spherical \cite{belv}. Thus, many theoretical studies have taken into account the rotation effect of rapidly rotating neutron stars (i.e., pulsars) \cite{kurk, orsa1, orsa2, webe}. In this paper, we study the EOS for a rotating star.

\section{Formalism and Numerical Results}
We use our previous formulation \cite{end2,end3}, which we briefly review here. The quark phase consists of the lighter {\it u}, {\it d}, and {\it s} quarks, together with electrons. We employ the MIT bag model and assume a sharp boundary at the quark--hadron interface: the {\it u} and {\it d} quarks are treated as massless and {\it s} as massive ($m_s=150$MeV), and they interact with each other via one-gluon-exchanges, which occur inside the bag. The hadron phase consists solely of protons, neutrons and electrons. We adopt an effective potential to reproduce the saturation properties of nuclear matter. We consider the phase transition at the interface between these two phases. When treating the phase transition, we have to consider thermodynamic potentials of each constituent. Then the total thermodynamic potential ($\Omega_\mathrm{total}$) consists of the hadron, quark, and electron contributions as well as a surface contribution:
\begin{equation}
\Omega_\mathrm{total} = \Omega_\mathrm{hadron} +\Omega_\mathrm{quark}
 +\Omega_\mathrm{surface}. 
\label{ometot}
\end{equation}
The surface contribution $\Omega_\mathrm{surface}$ is parameterized by the surface tension parameter $\sigma$, $\Omega_\mathrm{surface} = \sigma S$ with $S$ being the area of the interface. Note that this parameter may be closely related to the confining mechanism, but unfortunately we have no definite idea about how to incorporate it into that mechanism. Indeed, many authors treat the strength of $\sigma$ as a free parameter and observe the variation in the results with different settings\cite{pet,gle2,alf2}. Taking the same approach for the surface contribution,  we carefully minimize $\Omega_{\rm total}$ in this study, with the Gibbs conditions for phase equilibrium:
\begin{equation}
 \mu_{\mathrm{B}}^{\mathrm{Q}} = \mu_{\mathrm{B}}^{\mathrm{H}} (\equiv \mu_{\mathrm{B}} ),
 \hspace{5pt} \mu_{\mathrm{charge}}^{\mathrm{Q}} =
 \mu_{\mathrm{charge}}^{\mathrm{H}}, \hspace{5pt} P^{\mathrm{Q}}=P^{\mathrm{H}}, \hspace{5pt} T^{\mathrm{Q}} = T^{\mathrm{H}},
\label{gc}
\end{equation}
where the superscript H(Q) denotes the hadron (quark) phase, and $\mu_{\mathrm{B}}^\mathrm{H(Q)}$ and $\mu_{\mathrm{charge}}^\mathrm{H(Q)}$ are the baryon number and charge chemical potentials, respectively. Note that there are two independent chemical potentials in this phase transition.
In this case, the transition should be much different from the liquid--vapor phase transition, which is described by a single chemical potential.

To clarify charge-screening effects, we also perform calculations without including screening \cite{end2, maru, maru2}. We also adopt the EOS derived in a previous paper \cite{end2} and then apply it to the TOV equation \cite{end3,end4}. For comparison, we perform a calculation for a stationary rotating star using our EOS.

However, taking rotational effects into account in general relativity is notoriously difficult. Therefore, we use the following assumptions regarding the star's rotation and state: 
1. Stationary rigid rotation (``uniform rotation''); 
2. Axial symmetric rotation with respect to the spin axis; and 
3. The matter is a perfect fluid.
Stationary rotation in general relativity has been reviewed in \cite{ster} as well as in \cite{kurk}; we follow their calculations here in applying our EOS to the stationary rotating star.
\begin{figure}[htb]
\begin{center}
\includegraphics[width=75mm]{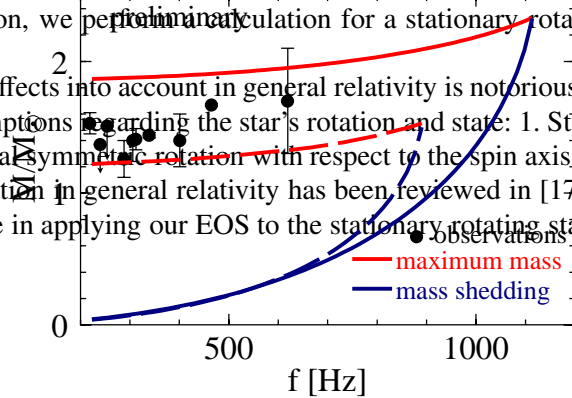}
\caption{(Color online) Mass--frequency relation for our models with (solid curve) and without (dashed curve) screening plotted with the observational data \cite{kurk}.}
\label{M-f}
\end{center}
\end{figure}

Figure\ \ref{M-f} shows the results for a rotating star using our EOSs with and without screening. The red curve corresponds to the maximum mass of the star, while the blue curve results from mass shedding and corresponds to the ``Kepler frequency". The Kepler frequency indicates that the centrifugal force equals the force of gravity. Therefore, the region to the right of the blue curve is physically invalid. If the red curve is below the plotted observational data, the EOS must be ruled out. Hence our EOS result with screening is consistent with these observations whereas that without screening is not and therefore unsuitable. To confirm this possibility, we need to conduct further studies, the behavior may be attributable to the softness of the EOS\cite{kurk}.

\section{Summary}
We have revealed the differences between applying screening and no-screening when taking into account rotation effects on quark--hadron hybrid stars. We adopted simple assumptions in modeling both quark and nuclear matter. To obtain a more realistic picture of the quark--hadron phase transition, we need to take into account color superconductivity for quark matter \cite{alf1, alf2, ippo, ayva} and relativistic mean field theory for nuclear matter \cite{shen}. We should then be able to provide more realistic results. Neutron stars have other important features, such as magnetic fields \cite{dex1, chir}. While the origin of these fields is still unknown, there are possible ways to explain it based on the spin polarization of the quark matter \cite{tat1, tat3, tat4, tsue1, tsue2}. However, whether quark matter exists strongly depends on the EOS. In this calculation, we did not take into account such magnetic fields in rapidly rotating neutron stars. If we consider an magnetic field, we may be able to elucidate the EOS for high--density quark--hadron matter.

\end{document}